\documentstyle[aps,preprint,tighten]{revtex} 
\begin{document}
\draft
\preprint{imsc/98/04/16}
\title{General solution of the non-abelian Gauss law and non-abelian 
analogs of the Hodge decomposition} 
\author{Pushan Majumdar \thanks{e-mail:pushan@imsc.ernet.in}
\and H.S.Sharatchandra \thanks{e-mail:sharat@imsc.ernet.in}} 
\address{Institute of Mathematical Sciences,C.I.T campus Taramani.  
Madras 600-113} 
\maketitle
\begin{abstract} 
General solution of the non-abelian Gauss law in terms of covariant 
curls and gradients is presented. Also two non-abelian analogs of the 
Hodge decomposition in three dimensions are addressed. i) Decomposition 
of an isotriplet vector field $V_{i}^{a}(x)$ as sum of covariant curl 
and gradient with respect to an arbitrary background Yang-Mills 
potential is obtained. ii) A decomposition of the form 
$V_{i}^{a}=B_{i}^{a}(C)+D_{i}(C) \phi^{a} $ 
which involves non-abelian magnetic field 
of a new Yang-Mills potential C is also presented. These results are 
relevant for duality transformation for non-abelian gauge fields.
\end{abstract} 

Yang-Mills theory has the conjugate variables $\vec{A}_{i}(x)$ and 
$\vec{E}_{i}(x)$, where $\vec{A}_{i}(x)=A_{i}^{a}(x), (i,a=1,2,3)$. 
$\vec{A}_{i}(x)$ is the Yang-Mills potential 
and $\vec{E}_{i}(x)$ is the non-abelian electric field. There is a first 
class constraint, the Gauss law,
\begin{equation}\label{gauss}
\partial_{i}\vec{E}_{i} + \vec{A}_{i}\times \vec{E}_{i}=0.
\end{equation}
This particular constraint is also present in Ashtekar formulation of 
gravity. Usually this constraint is handled by ``fixing a gauge". 
However it is of interest to obtain a parametrization of the ``physical 
phase space", i.e. the part of the phase space which satisfies the 
constraint (\ref{gauss}). This would give the physical degrees of 
freedom. For other approaches to handle the Gauss law, see ref. 
\cite{previous}. We are interested in a general solution of 
(\ref{gauss}) in terms of local fields. This is of relevance for duality 
transformation of Yang-Mills theory \cite{dual}\cite{Haag}\cite{Puri}. Note 
that in the abelian 
case the Gauss law constraint $\partial_{i}E_{i}=0$ has the solution
$E_{i}=\epsilon_{ijk}\partial_{j}C_{k}$. Here $C_{k}$ turns out to be 
the dual gauge potential which couples minimally to magnetic matter.

In analogy to the abelian case, we consider the ansatz
\begin{equation}\label{ansatz}
E_{i}=\epsilon_{ijk}[D_{j}[A],C_{k}].
\end{equation}
Here we are using the matrix notation; 
$A_{i}=\vec{A}_{i}\cdot\vec{\sigma}/2$ etc., where $\sigma^{a}$ are the 
Pauli matrices and $D_{j}[A]={\bf 1}\partial_{j}+A_{i}$.
Substituting in (\ref{gauss}) and using the Jacobi identity, we get,
\begin{equation}\label{ag}
[B_{i}[A],C_{i}]=0
\end{equation}
where sum over $i$ is implied. Here
\begin{equation}\label{defB}
\vec{B}_{i}[A](x)=\epsilon_{ijk}(\partial_{j}\vec{A}_{k}+
\frac{1}{2} \vec{A}_{j}\times \vec{A}_{k})
\end{equation}
is the non-abelian magnetic field. We consider a generic case 
where the $3\times 3$ matrix $B_{i}^{a}$ is invertible in a certain 
region of space. Then we may use $B_{i}^{a}$ to ``lower" the index $a$ 
in $C_{i}^{a}$:
\begin{equation}\label{lower}
C_{i}^{a}=C_{ij}B_{j}^{a}.
\end{equation}
From (\ref{ag}), we get, $|B|(B^{-1})_{a}^{k}(\epsilon_{ijk}C_{ij})=0$, 
where $|B|=det(B_{i}^{a}).$
Therefore equation (\ref{ag}) is satisfied if and only if $C_{ij}$ is an 
arbitrary symmetric matrix. Thus we have obtained a class of solutions 
$E_{i}=\epsilon_{ijk}(B_{l}[A]\partial_{j}C_{lk} 
+[D_{j}[A],B_{l}[A]]C_{lk}.$ This presents the 
solution as a covariant curl in close analogy to the abelian case. 

The symmetric tensor field $C_{ij}$ has six degrees of freedom at each $x$. 
Therefore it would appear that the solution in terms of the gauge invariant 
field $C_{ij}$ is a general solution. We now argue that this is not the 
case.

Two fields $C$ and $C^{\prime}$ give the same solution of the 
non-abelian Gauss law if 
\begin{equation}\label{cop}
\epsilon_{ijk}[D_{j}[A],e_{k}]=0,
\end{equation}
where $e_{k}=(C-C^{\prime})_{k}.$ This is precisely the equation for a 
driebein $e_{i}^{a}$ to be torsion free with respect to the connection 
one form $\omega_{i}^{ab}=\epsilon^{abc}A_{i}^{c}.$ This situation has 
been analyzed in detail in \cite{gfc}. For any $A_{i}^{a}$ there are 
solutions 
$e_{i}^{a}[A,\theta]$ labeled by an arbitrary triplet of fields 
$\vec{\theta}(x)$. This means that our ansatz (\ref{ansatz}) gives 
solutions labeled by only three degrees of freedom at each $x$.

In order to obtain a general solution of equation (\ref{gauss}), we 
include a covariant gradient also in the ansatz:
\begin{equation}\label{newan}
E_{i}=\epsilon_{ijk}[D_{j}[A],C_{k}] + [D_{i}[A],\phi].
\end{equation}
This satisfies equation (\ref{gauss}) if
\begin{equation}\label{nsol}
[B_{i}[A],C_{i}]+[D_{i}[A],[D_{i}[A],\phi]]=0
\end{equation}
Using (\ref{lower}) this may be rewritten as
\begin{equation}\label{asym}
\epsilon_{ijk}C_{ij}=-\frac{1}{|B|} 
\vec{B}_{k}\cdot D^{2}[A]\vec{\phi}. 
\end{equation}
For any $\phi$, this locally determines the antisymmetric part of $C_{ij}$.
Thus we get a general solution parametrized by an arbitrary symmetric 
tensor field $C_{ij}$ and an isotriplet scalar field $\phi^{a}(x)$.
Alternately we may choose $C_{i}^{a}$ arbitrary and adjust $\phi$ to satisfy
(\ref{asym}).

In contrast to the abelian case, the (covariant) gradient necessarily 
appears in the general solution, though it is determined by the 
(covariant) curl part. Here $C$ and $C+e[A,\theta]$ reproduce 
the same solution. This is the analog of the fact that both $C_{i}$ and 
$C_{i}+\partial_{i}\theta$ give the solution of the Gauss law in Maxwell 
theory. Our solution (\ref{newan}) is the analog of the Poincare lemma 
for the non-abelian case.

Thus the gauge invariant second rank tensor $C_{ij}$, in equation 
(\ref{lower}), (modulo the invariance $C\rightarrow C+e[A,\theta]$) 
completely describes the physical degrees of freedom of the Yang-Mills 
theory.

Given an $E_{i}$ satisfying the non-abelian Gauss law (\ref{gauss}), 
construction of $C$ and $\phi$ is as follows. Applying covariant curl on 
both sides of (\ref{newan}), we get,
\begin{equation}\label{recon}
\epsilon_{ijk}[D_{j}[A],E_{k}]= -[D_{j}[A],[D_{j}[A],C_{i}]]
+[D_{j}[A],[D_{i}[A],C_{j}]]+[B_{i}[A],\phi]
\end{equation}
Now from (\ref{nsol})
\begin{equation}\label{phi}
\phi=-D^{-2}[A][B_{j},C_{j}].
\end{equation}
Substituting this $\phi$ in the previous equation, we get a second order 
equation for $C$ in terms of $E$. The solution $C$ is of course not 
unique because $C$ and $C+e[A,\theta]$, both give the same $E$. In other 
words the second order operator acting on $C$ has many null eigen-vectors.
It has been argued in \cite{gfc}, that by requiring $C$ to satisfy a 
``gauge 
condition" such as $\partial_{i}C_{i}=0$ or $D_{i}C_{i}=0$, we get a 
unique solution for $C$. Thus $C$ maybe computed and also $\phi$ using
(\ref{phi}).

We may use the above results for obtaining the non-abelian analogs of 
the Hodge decomposition. Consider any isotriplet vector field 
$V_{i}^{a}(x)$. We first consider a decomposition of $V_{i}$ as a sum 
of a covariant curl and a covariant gradient with respect to any 
specified Yang-Mills potential $A_{i}^{a}(x)$. Consider
$\vec{\cal E}_{i}=\vec{V}_{i}-D_{i}[A]D^{-2}[A](D_{j}[A]\vec{V}_{j}).$ 
Here we are presuming that the covariant laplacian $D^{2}[A]$ has no 
zero eigenvalues and is therefore invertible. This would be true for 
fields vanishing rapidly at infinity on ${\bf R}^{3}$. Thus any $V_{i}^{a}$
has a unique decomposition $\vec{V}_{i}=D_{i}[A]\vec{\chi}+\vec{\cal E}_{i}$
where $D_{i}\vec{\cal E}_{i}=0$. For ${\cal E}_{i}$ we have a general 
decomposition as in (\ref{newan}). Thus we have a decomposition of 
$V_{i}$ into covariant curl and covariant gradient.

The above procedure may also be generalized when the covariant laplacian 
$D^{2}[A]$ has null eigen-vectors, for example on compact manifolds. In 
this case a ``harmonic form" is also required for the decomposition. We 
may expect this harmonic part to have a cohomological interpretation.

Next we consider a different non-abelian analog of the Hodge 
decomposition. We seek a decomposition,
\begin{equation}\label{decomp}
\vec{V}_{i}=\vec{B}_{i}[C]+D_{i}[C]\vec{\phi}
\end{equation}
in terms of the non-abelian magnetic field and covariant gradient with 
respect to a new gauge potential $C$. Note that in contrast to the 
previous case, $B_{i}[C]$ is non-linear in $C$. Therefore even if the 
decomposition exists, the reconstruction of $C$ is not easy. In case a 
specified background field $A$ is ``close" to $C$,
\begin{equation}\label{approx}
\vec{V}_{i}-\vec{B}_{i}[A]\simeq \epsilon_{ijk}D_{j}[A]\Delta 
\vec{C}_{k}+D_{i}[A]\vec{\phi}
\end{equation}
where $\Delta \vec{C}_{k}=\vec{C}_{k}-\vec{A}_{k}.$ Thus the previous 
decomposition may be 
regarded as a special case of this when the given vector field $V_{i}$ is  
``close" to $B_{i}[A]$ for the specified Yang-Mills potential $A_{i}$.

We first note that $B_{i}[C]$ and $D_{i}[C]\phi$ represent independent 
degrees of freedom of $V_{i}$ just as curl and gradient. As a 
consequence of the bianchi identity, the inner product $\int d^{3}x 
\vec{B}_{i}[C]\cdot D_{i}[C]\vec{\phi}=\int 
dS_{i}\:\vec{\phi}\cdot\vec{B}_{i}$ which gives the first Chern class. 
For fields vanishing rapidly at infinity, this is zero. Moreover the 
equation $\vec{B}_{i}[C]=D_{i}[C]\vec{\phi}$ is precisely the Bogomolnyi 
equation. All solutions of this are known by ADHM construction and are 
labeled by positions and (isospin) orientations of (anti-)monopoles. In 
case we 
require fields vanish faster than $r^{-1}$ at infinity, then there are 
effectively no solutions. Thus $B_{i}[C]$ and $D_{i}[C]\phi$ represent 
distinct degrees of freedom.

In order to construct $C$ and $\phi$, we consider an interpolation 
procedure. Consider
\begin{equation}\label{inter}
\lambda \vec{V}_{i}
=\vec{B}_{i}[C(\lambda)]+D_{i}[C(\lambda)]\vec{\phi}(\lambda)
\end{equation}
where 
\begin{equation}\label{ser}
C(\lambda)=\sum_{n=1}^{\infty} \lambda^{n}C^{(n)}
\;{\rm and}\;
\phi(\lambda)=\sum_{n=1}^{\infty} \lambda^{n}\phi^{(n)}
\end{equation}
Terms linear in $\lambda$ give
$\vec{V}_{i}=\epsilon_{ijk}\partial_{j}\vec{C}_{k}^{(1)}+\partial_{i} 
\vec{\phi}^{(1)}$, which is just the usual Hodge decomposition of $V_{i}$.
We know the decomposition exists and is unique. $C_{k}^{(1)}$ and 
$C_{k}^{(1)}+\partial_{k}\Lambda$, both give the same decomposition. Terms 
quadratic in $\lambda$ gives $\epsilon_{ijk}\partial_{j}\vec{C}_{k}^{(2)}+ 
\partial_{i}\vec{\phi}^{(2)}=-\epsilon_{ijk}\vec{C}_{j}^{(1)}\times
\vec{C}_{k}^{(1)}-\vec{C}_{i}^{(1)}\times\vec{\phi}^{(1)}.$ 
$C^{(1)}$ and $\phi^{(1)}$ are already known. Hence $\phi^{(2)}$ and 
$C^{(2)}$ 
are determined. Again the gradient part of $C^{(2)}$ is arbitrary. This 
way all the $C^{(n)}$'s and $\phi^{(n)}$'s are determined successively. 
If we 
impose a ``gauge condition" such as $\partial_{i}\vec{C}_{i}^{(n)}=0$, 
$C^{(n)}$ and $\phi^{(n)}$ are unique at each stage.
This interpolation procedure makes the connection to the usual Hodge 
decomposition explicit and provides a plausible technique for 
reconstucting $C$ and $\phi$. We do not address the question of 
convergence in (\ref{ser}), but provide an alternate procedure for 
reconstruction of $C$ and $\phi$ below.

A way of avoiding interpolation is as follows. Consider,
\begin{equation}\label{alt}
\vec{\cal E}_{i}[C]=\vec{V}_{i}-D_{i}[C]D^{-2}[C]D_{j}[C]\vec{V}_{j}
\end{equation}
which satisfies $D_{i}[C]\vec{\cal E}_{i}[C]=0.$ We have to choose 
$C$, such that
\begin{equation}\label{trans}
\vec{e}_{i}[C,\theta] \times \vec{\cal E}_{i}[C]=0
\end{equation}
for every driebein $e_{i}[C,\theta]$ which is torsion free with respect 
to the connection one form $C$. Then, ${\cal E}_{i}[C]$ is the 
non-abelian 
magnetic field $B_{i}[C]$ and we have the decomposition (\ref{decomp}).

It is clear that the gradient part is present for an arbitrary $V_{i}$. 
$B_{i}[C]$ provides only six degrees of freedom at each $x$ and $\phi$ 
the other three. Thus it is not possible to write an arbitrary $V_{i}$ 
as a non-abelian magnetic field. The conditions that $V_{i}$ has to 
satisfy in order to be a non-abelian magnetic field are presented in 
\cite{gfc}.

We have obtained a general solution of the non-abelian Gauss law in 
close analogy to the Poincare lemma. We have used it to address the 
non-abelian analog of the Hodge decomposition. These are useful for 
duality transformation of Yang-Mills theory 
\cite{dual}\cite{Haag}\cite{Puri}.

One of us (H.S.S.) thanks Prof. Ramesh Anishetty and Dr.G.H.Gadyar for some 
important remarks.

\end{document}